\documentclass[aps,prl,showpacs,reprint,longbibliography]{revtex4-1}
\usepackage{bm,epsfig,graphics,amssymb,amsmath,subeqnarray,setspace,graphicx,amsthm,epstopdf,subfigure,color,txfonts}

\usepackage{setspace,natbib}

\usepackage{epsfig,graphics,amssymb,amsmath,color,setspace}
\usepackage{graphicx}
\usepackage{natbib}
\usepackage[sfdefault=cmbr]{isomath}
\usepackage{tikz}

\usepackage{graphicx,color,bm}
\graphicspath{{converted_graphics/}}
\usepackage{graphicx}


\begin{document}

\title{Force induced formation of twisted chiral ribbons}
\author{Andrew Balchunas$^1$, Leroy L. Jia$^2$,  Mark Zakhary$^1$, Zvonimir Dogic$^{1,3}$, Robert A. Pelcovits$^4$, and Thomas R. Powers$^{4,5}$}
\affiliation{$^1$The Martin Fisher School of Physics, Brandeis University, 415 South St, Waltham, MA 02454, USA}
\affiliation{$^2$Division of Applied Mathematics, Brown University, Providence, RI 02912, USA}
\affiliation{$^3$Department of Physics, University of California at Santa Barbara, Santa Barbara, CA 93106,}
\affiliation{$^4$Department of Physics, Brown University, 182 Hope Street, Providence, RI 02912, USA}
\affiliation{$^5$School of Engineering, Brown University, 182 Hope Street, Providence, RI 02912, USA}

\date{\today}

\begin{abstract} 
We study the emergence of helical structures subjected to a stretching force, demonstrating that the force transforms disk-shaped colloidal membranes into twisted chiral ribbons of predetermined handedness. Using an experimental technique that enforces torque-free boundary conditions we simultaneously measure the force-extension curve and quantify the shape of emergent ribbons. An effective theory that accounts for the membrane bending energy and uses geometric properties of the edge to model the internal liquid crystalline degrees of freedom explains both the measured force-extension curve and shape of the twisted ribbons in response to an applied force.
\end{abstract}

\maketitle

Helical shapes are ubiquitous in Nature, arising at all length scales ranging from microscopic assemblages such as collagen, bacterial flagella and lipid bilayer membranes to macroscopic seed pods and the helical structures of the climbing plants~\cite{darwin1875movements,berg2003rotary,burkhard2001coiled,brodsky1997collagen,schnur1993lipid,oda1999tuning,studart2014bioinspired,armon2011geometry,klein2007shaping}. Chiral superstructures are also found in diverse synthetic materials including nanoparticle based photovoltaic assemblages and textiles~\cite{lima2011biscrolling,sone2002semiconductor,srivastava2010light}. How the molecular chirality of the microscopic constituents influences the emergent chiral shape is an important question that underlies all of these phenomena. Generally, an external force unwinds a helix thus reducing its chirality~\cite{cluzel1996dna,smith1996overstretching,smith2001tension}. Here, we study the emergence of chiral shapes in 2D  colloidal  membranes, which are one-rod-length-thick liquid-like monolayers of aligned rods that in the presence of non-adsorbing polymer assemble into diverse structures~{\cite{barry2010entropy,yang2012self,gibaud2012reconfigurable,zakhary2014imprintable,sharma2014hierarchical,gibaud2017achiral}.  Colloidal membranes also form twisted ribbons which requires coupling of liquid crystalline order of the constituent chiral rods with the shape of the membrane's edge~\cite{kaplan2010theory,tu2013theory,jia2017chiral}. A quantitative understanding of twisted colloidal ribbons and their mechanical properties might be relevant for understanding the generic pathways by which microscopic chirality is expressed on the macroscopic scales. 

In our experiments we assemble rod-like viruses into equilibrium flat circular disks that lack any apparent helicity. Surprisingly, we find that application of a sufficiently strong force transforms such achiral assemblages into chiral twisted ribbons of predetermined handedness. This observation provide a unique opportunity to study the emergence of force induced chiral structures for several reasons. First, colloidal membranes have a vanishing zero-frequency in-plane shear modulus. Consequently, in contrast to solid elastic sheets, colloidal membranes easily reconfigure, switch between different topologies and change Gaussian curvature when subjected to external forces. Second, extensive previous work has demonstrated that shapes of colloidal membranes can be described with an effective elastic theory in which the bending deformations are described by the Helfrich-Canham free energy~\cite{canham1970minimum,helfrich1973elastic}, while liquid-crystalline degrees of freedom of edge-bound rods are described by geometric quantities such as the length, curvature, and geodesic torsion~\cite{gibaud2012reconfigurable,barry2008direct,jia2017chiral}. Most of the properties that govern these deformation modes have been measured independently thus allowing for rigorous parameter-free tests of theoretical models. For example, colloidal membranes have an intrinsic preference for surfaces with negative Gaussian curvature~\cite{gibaud2017achiral,jia2017chiral}. The Gaussian modulus, $\bar\kappa$ that controls this preference is important in colloidal and lipid membranes alike, but due to their larger size, slower time scale for shape changes and tendency to form structures with open edges, it is easier to measure $\bar\kappa$ for colloidal membranes. These features enable us to develop a quantitative model, without any adjustable parameters, that predicts the experimentally observed shape and twist of the colloidal ribbons as a function of applied force.

\begin{figure}[h]
\centering
\includegraphics[scale=0.4]{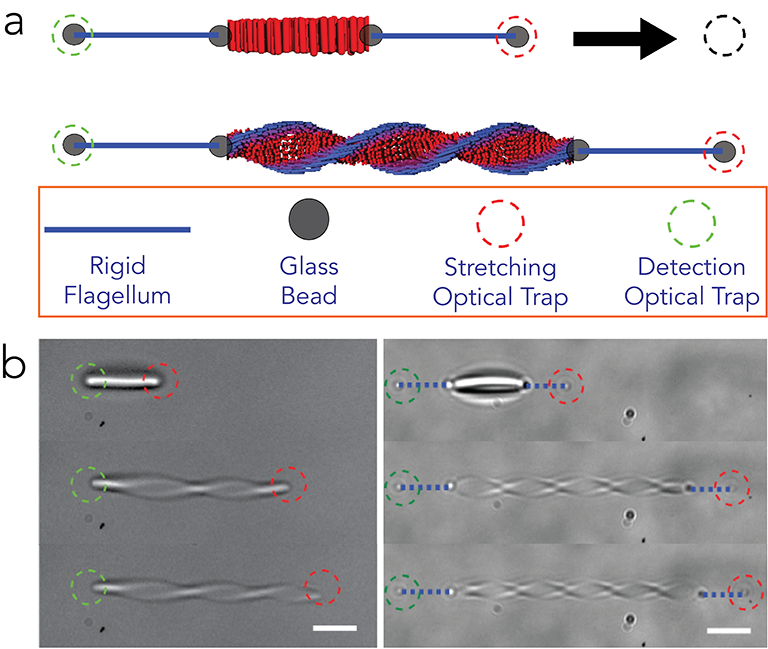}
\caption{(Color online.) Stretching a colloidal membrane induces twist. a) Schematic of a setup that applies a stretching force with a torque-free boundary condition. Side view of flagellar dumbbells attached to an unstretched membrane (top) and a twisted ribbon (bottom). (b) Left: Directly trapping colloidal membranes generates an external torque. With increasing extension the ribbon twists when the internal torque generated by the twist overcomes the external torque induced by the optical trap. Right: A membrane stretched with flagellar dumbbells twists continuously since the optical trap is removed from the membrane and symmetric trapped beads freely rotate. Dashed circles indicate the positions of optical traps. Scale bars, 5 $\mu$m.}\label{expsetup}
\end{figure}

We studied membranes assembled from either 0.88 $\mu$m long {\it fd-wt} or 1.2 $\mu$m long M13KO7 viruses that also differ in surface  charge~\cite{dogic2001development,grelet2003origin}. Our experimental setup allowed us to simultaneously apply a known force on the membrane and observe its shape (Fig.~\ref{expsetup}a)~\cite{siavashpouri2017molecular}. Two optical traps are produced using an acousto-optical deflector. One trap held the membrane fixed while the other trap placed on the opposite side extended the membrane with well defined steps. Directly trapping the colloidal membrane exerted both a torque and an extension force that precluded accurate measurements (Fig.~\ref{expsetup}b). To generate torque free boundary conditions we assembled ``flagella dumbbells" by binding two streptavidin coated silica beads ($2\,\mu$m diameter) to both ends of a rigid biotin labelled straight flagellar filament isolated from strain SJW 1660 (Fig.~\ref{expsetup}b)~\cite{memet2018microtubules}. The beads were also coated with an antibody for filamentous viruses which induced strong biding to the membrane edge. Experiments were performed in a microfluidic T-chamber where a membrane forming suspension of viruses (100 mM or 125 mM NaCl and 20 mM tris, pH=8.15) and Dextran (M.W 500,000) was flowed into the vertical stem, and a suspension of biotinylated flagellar filaments, Dextran and streptavidin and antibody coated silica beads in the same buffer were flowed into the perpendicular arm of the channel. The sample was prepared 24 hours before experiments and kept hydrated to allow colloidal membranes to assemble. Two flagella dumbbell handles were first constructed in the vertical arm using steerable optical traps. Subsequently, using the same optical traps the ``flagella dumbbell'' handles were moved to the center of the T-stem. One bead of each dumbbell was attached to the opposite sides of an isolated membrane, and the free beads were optically trapped.  The flagellar dumbbells accurately propagate the applied force due to their high stiffness (a persistence length of $\sim1$ mm)~\cite{memet2018microtubules}.

\begin{figure}[t]
\centering
\includegraphics[scale=0.55]{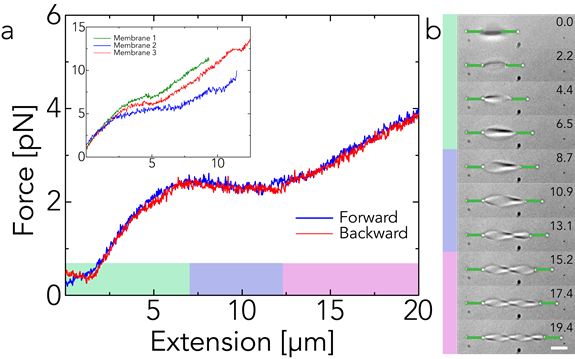}
\caption{(Color online.) (a) Measurement of the force-extension curve for a single $fd$-$wt$ membrane (6.4 mm diameter, 45 mg/ml Dextran and 125 mM NaCl). At low extensions (green bar) the force is directly proportional to extension. At intermediate forces one observes a force plateau (blue bar), and then again a linear increase in force (magenta bar). Inset: Force-extension curves for three different membranes of comparable size ({$\sim$6.2 $\mu$m}, 50 mg/ml Dextran, 100 mM NaCl) showing sample to sample measurement variation. (b) Shapes of twisted membrane corresponding to the measured force-extension curve. The extension is labeled in microns. Scale bar, 5\,$\mu$m.}\label{shapefig}
\label{regimes}
\end{figure}

Using the above described setup we characterized the response of colloidal membranes to applied force. For any given membrane the measurement was highly repeatable, and there was no hysteresis; we measured the same curve with increasing or decreasing extension (Fig.~\ref{shapefig}a). However,  for different membranes of comparable diameter, measurements had ~20\% variability in the force extension curve (Fig.~\ref{shapefig}a, inset), which is presumably due to variations in  dumbbells attachments to the membrane. In general, the measured force-extension curve exhibited three regimes. For small extensions, the force increased linearly with extension. At intermediate forces the force plateaued even as the extension changed by up to a few hundred percent.  At even higher extension the plateau regime transitioned to another ``overstretched'' linear regime (Fig.~\ref{shapefig}a). Our setup allowed us to observe the membrane shape while measuring the force-extension curve (Fig.~\ref{shapefig}b). For smallest extensions in the linear regime, the disk-shaped membrane elongated, but exhibited no measurable twist. As the extension increased, the membrane started to twist, with the twist appearing gradually rather than abruptly, and the handedness of the twist was always the same. The onset of measurable twist roughly coincided with the appearance of the force plateau. 

We measured how the force-extension curve depends on the diameter for membranes assembled from M13-KO7 virus that have an edge tension of  $\sim1000\,k_\mathrm{B}T/\mu\mathrm{m}$ (Fig.~\ref{M13fig}a). The initial slope in the weak force extension regime increased with decreasing membrane size (Fig.~\ref{M13fig}b). Furthermore, the transition from the plateau to the overstretched regime occurred at lower extensions for smaller diameter membranes. Notably, for the smallest diameter membranes, we observed no discernible force plateau as the linear and ``overstretched'' regimes effectively merged. We have also examined how the force-extension curve depends on the membrane edge energy. To accomplish this we studied colloidal membranes assembled from {\it fd-wt} which had a line tension of $\sim380\,k_\mathrm{B}T/\mu \mathrm{m}$ (Fig.~\ref{fig4}). The magnitude of the force plateau was dependent on the edge tension, being 2--3\,pN for the low tension {\it fd-wt} membranes (Fig.~\ref{fig4}), and 6--10\,pN for the high tension M13-KO7 membranes (Fig.~\ref{M13fig}a). Finally, our setup allowed us to quantify the spontaneous twist as a function of the extension (Fig.~\ref{fig4} inset). Since the pitch is defined to be twice the distance between two nodes of the twisted ribbon, it was difficult to measure the pitch at small extensions when there is one or no node. 
 
To understand how extension induces twist we use an effective model of colloidal membranes which assumes that membrane thickness is small compared to its radius of curvature~\cite{jia2017chiral}. In this limit the surface deformation is described by the Canham-Helfrich bending energy:
\begin{equation}
E_{\mathrm{CH}}=\frac{\kappa}{2}\int dA (2H)^2 + \bar\kappa \int dA K,
\end{equation}
where $H$ is the mean curvature and $K$ is the Gaussian curvature~\cite{canham1970minimum,helfrich1973elastic}. Recent estimates of the elastic moduli $\kappa$ and $\bar\kappa$ for colloidal membranes suggest $\kappa\gg\bar\kappa$, in contrast to  surfactants  or lipid bilayers, for which $\kappa$ and $\bar\kappa$ are typically comparable~\cite{jung2002gaussian, hu2012determining}. Experiments and a simple theoretical model estimate $\bar\kappa\approx200\,k_\mathrm{B}T$~\cite{gibaud2017achiral,jia2017chiral}. Measurements of the areal compressibility modulus of colloidal membranes yielded a bending stiffness of $\kappa \approx 15,000\,k_BT$. The large value of $\kappa$ implies that colloidal membranes ordinarily form minimal or nearly minimal surfaces, with negative Gaussian curvature which are compatible with a positive value of $\bar\kappa$. 

Unlike a simple interface, the membrane edge also has a considerable bending rigidity, which is $\sim 100\,k_\mathrm{B}T \mu m$ for both membrane types. This bending rigidity stems from the configuration of the rods near the edge, which twist away from the membrane normal over the characteristic twist penetration length scale, and form a one-dimensional nematic phase that resists bending~\cite{barry2008direct,gibaud2012reconfigurable}. If the twist penetration length is small compared to the membrane size, we can account for the rod degrees of freedom with an effective energy that depends on geometric properties of the edge:
\begin{equation}
E_{\mathrm{edge}} = \int ds \left[\gamma+ \frac{B}{2}k^2 + \frac{B'}{2}\left(\tau_g-\tau_g^*\right)^2 \right],
\end{equation}
where $s$ and $k$ are the edge arclength and curvature, $\gamma$ is the line tension, $B$ is the effective edge bending stiffness, $\tau_g$  and  $\tau_g^*$ are the geodesic and  spontaneous geodesic torsion respectively, and $B'$ is the effective edge torsional stiffness~\cite{jia2017chiral}. $\tau_g^*$ is proportional to the rate that the rods twist in the cholesteric phase. We assume a chiral modulus $c^*=-B'\tau_g^*$, and $B'=B$. 

Because $\kappa\gg\bar\kappa$, the membrane takes the shape of a helicoid with a possibly infinite pitch $p$. 
The Gauss-Bonnet theorem transforms the integral of the Gaussian curvature over the surface into an integral of the geodesic curvature $k_g$ over the edge (up to a topological constant which we ignore). Therefore, the problem of finding the membrane shape for a given extension is reduced to finding the shape of the boundary of the membrane on a helicoid, where the pitch $p$ of the helicoid must also be determined. We limit our analysis to the initial linear and plateau regions of the force-extension curve. Subject to the constraint of the fixed area, our task is to find the closed curve on the helicoid and the helicoidal pitch which minimize the energy
\begin{equation}
E=\int ds \left(-\bar\kappa k_g + \gamma+ \frac{B}{2}k^2 + \frac{B}{2}\tau_g^2 + c^*\tau_g\right),\label{EcurveEqn}
\end{equation}
for a given extension.

The surface takes the form 
${\bf Y}(r,\theta) = (r\cos\theta, r\sin\theta, p\theta/2\pi)$, where $0\leq r \leq R(\theta)$ and $0\leq \theta \leq 2\pi Z/p$. Here, $Z=2a+z$ is the length of the stretched shape, $a$ is the initial radius of the disk, and $z$ is the extension. Note that the assumption of vanishing mean curvature implies that the pitch is uniform. We numerically solve for the shape of the boundary of the surface, which is the curve
$\mathbf{X}(\theta) = (R(\theta)\cos\theta, R(\theta)\sin\theta,c\theta)$. To simplify the analysis, we introduce $\psi$,  the angle between the tangent to the curve and the $r$-axis~\cite{seifert1991shape} when the shape is ``unwound'' into a planar shape. Note that the coordinates $r$ and $\psi$ and $z$ and $\psi$ are not independent; they are related by the constraint $dr/ds = \cos \psi$ and $dz/ds=-\sin\psi$, and care must be taken to properly enforce these constraints. The geometrical edge quantities are:
\begin{eqnarray}
k_n &=& -\dfrac{\sin(2\psi)p/(2\pi)}{p^2/(2\pi)^2+r^2}\\
k_g &=& \psi_s + \dfrac{r\sin\psi}{p^2/(2\pi)^2+r^2}\\
k^2 &=& k_n^2 + k_g^2\\
\tau_g &=& \dfrac{p/(2\pi)(1-2\cos^2\psi)}{p^2/(2\pi)^2+r^2}.
\end{eqnarray}
Once the shape is determined, the force as a function of $Z$ is given by
$F={dE}/{dZ}$. 

\begin{figure}[h!]
\includegraphics[scale=0.2]{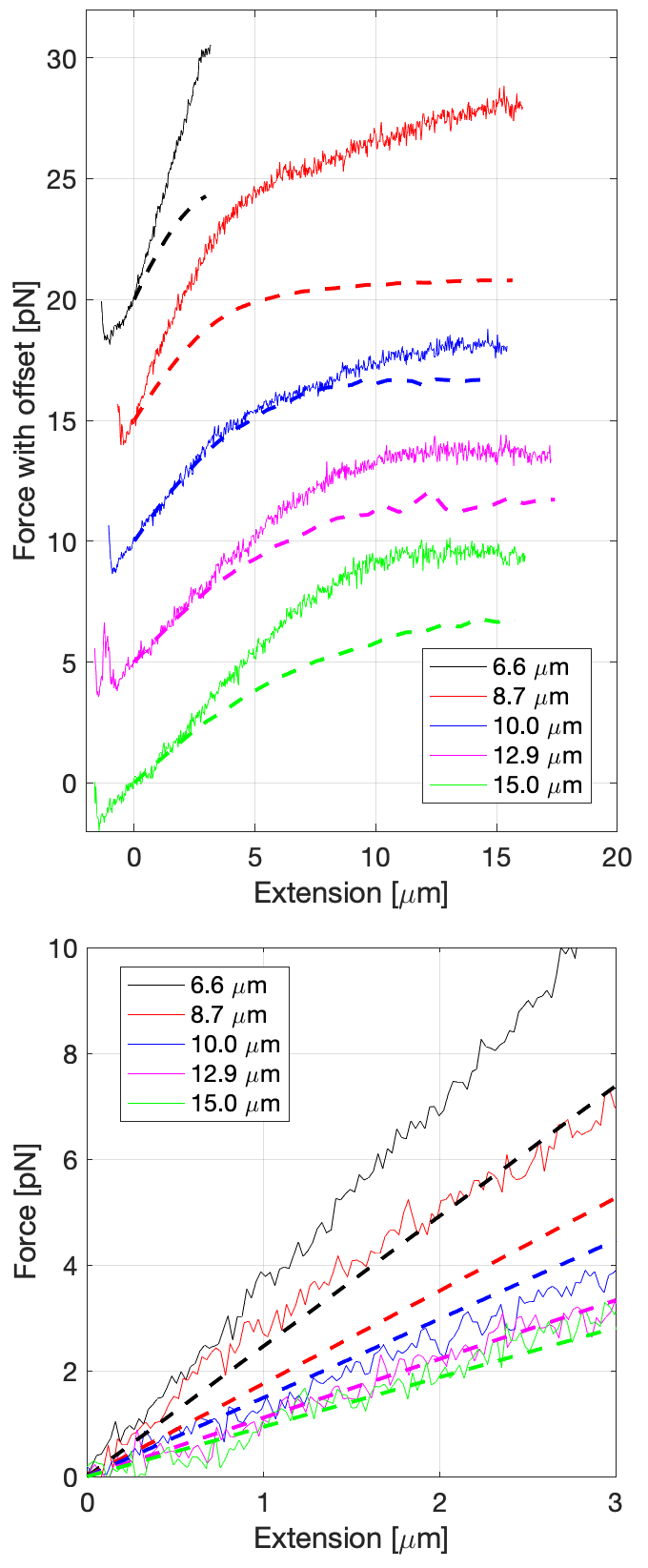}
\caption{Membrane diameter determines its stiffness. a) Experimental force-extension curves are shown with full lines for colloidal membranes of diameter that increases from 6.6\,$\mu$m to 15.0\,$\mu$m. The membranes are composed of $96\%$ M13-KO7 and $4\%$ M13 $wt$. For clarity, each membrane is offset by 5 pN. Theoretical predictions are shown with dashed lines using the following parameters: $\gamma=1000\,k_\mathrm{B}T/\mu$m~\cite{gibaud2012reconfigurable}, $\bar\kappa = 200\,k_\mathrm{B}T$~\cite{gibaud2017achiral}, $B=B'=100\,k_\mathrm{B}T \mu$m ~\cite{gibaud2012reconfigurable,jia2017chiral} and $c^* = 50\,k_\mathrm{B}T$ ~\cite{jia2017chiral}. b) Linear regime of the force-extension curves shown in panel a. Dashed lines are predictions of Eq.~\ref{kseq}. Membranes were assembled at 47\,mg\,mL$^{-1}$ Dextran, 100 mM NaCl.}
\label{M13fig}
\end{figure}

We first make analytical predictions about $F$ as a function of $z$. In the first linear regime $z\ll a$; we also assume $B\ll \gamma a^2$; disregarding the small twist, we find $F=k_\mathrm{s}z$, where the effective spring constant is:
\begin{equation}
k_\mathrm{s}=\frac{\pi \gamma}{2a} + \frac{\pi^2}{4}\sqrt{\frac{\gamma B}{a^4}}+\hdots.\label{kseq}
\end{equation}
In the applicable regime this approximation agrees with the full numerical force-extension described below, while qualitatively also reproducing experimental finding that the smaller membranes are effectively stiffer. Using independently determined values for $a$, $B$ and $\gamma$ we find that the above approximation is in quantitative agreement with the experiments, for membranes with diameter greater than 10\,$\mu$m (Fig.~\ref{M13fig}b). A nonzero value of the edge bending stiffness $B$ is essential for obtaining agreement with the experiment. As the membrane size becomes comparable to the twist penetration length, the plateau regime disappears and the theory underestimates the effective membrane stiffness. In the other limit of $z\gg a$, we approximate the membrane shape as a twisted rectangular strip. Minimizing the energy in the achiral case $c^*=0$ yields 
\begin{eqnarray}
F = 2\gamma - \frac{\bar\kappa^2}{B} + O\left(\frac{1}{z^2}\right).
\end{eqnarray}
The force saturates at a value independent of the membrane size with the line tension, Gaussian curvature modulus, and edge bending stiffness primarily responsible for the force in the linear regime and the plateau regime. Including the chiral coupling $c^*$ term yields a more complex form for the asymptotic force but a similar numerical value for the experimentally relevant parameters.

\begin{figure}[th!]
\includegraphics[scale=0.2]{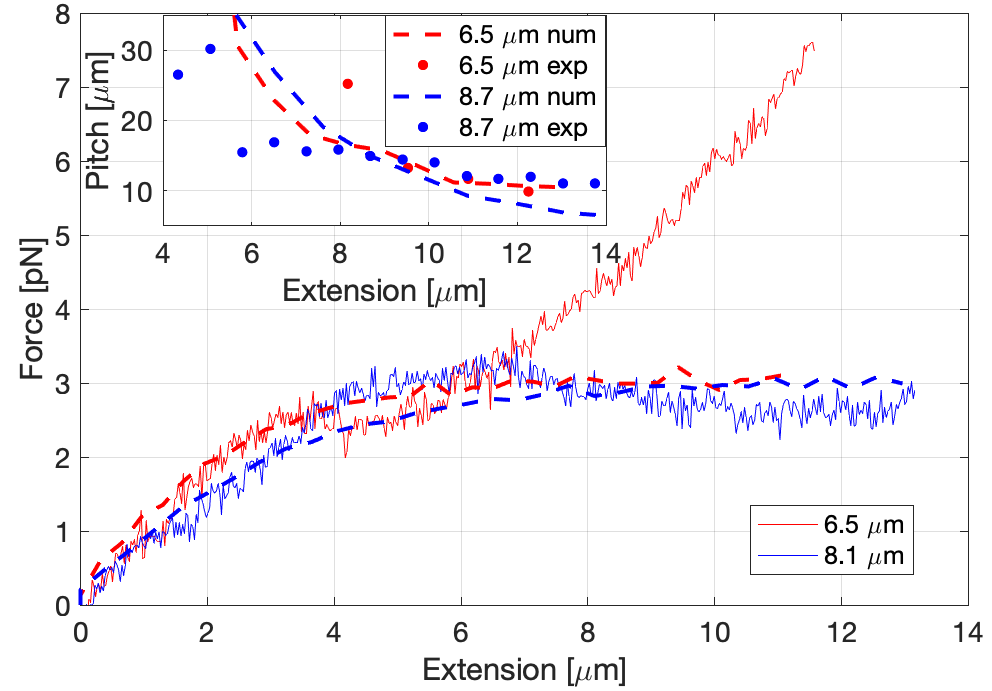}
\caption{ (Color online.)  Force-extension curves for $fd$-$wt$ membranes with $6.5\,\mu$m (red) and $8.1\,\mu$m (blue) diameters. Experimental data are indicated with full lines and numerical solutions with dashed lines. The line tension is 380 $k_BT$ $\mu$m. Other parameters used in the numerical calculation are the same as those for the M13-KO7 membranes. Dextran concentration is 50\,mg/mL and 100\,mM NaCl. Inset: Comparison between experimentally measured (circles) and numerically computed (dashed lines) pitch of the twisted ribbon as the function of the membrane extension.}
\label{fig4}
\end{figure}

Next we describe our numerical approach. To find the minimum of Eq.~\ref{EcurveEqn} we used a constrained interior-point optimization over discretizations of the variables $r(t)$ and $\psi(t)$ as well as the total arclength $L$ to calculate the continuous shape of minimal energy~\footnote{The MATLAB routine \texttt{fmincon} was used to perform the optimization}. Derivatives were approximated with second-order accurate central finite differences. We also enforced the global constraint of the constant area, $A=\pi a^2$, where $A$ is the total area of the helicoid and $a$ is the radius of the initial disk, as well as the compatibility condition relating $r$ and $\psi$ and $z$ and $\psi$ mentioned above. At the boundaries, we define the dimensionless arclength $t=s/L$, where $L$ is the (unknown) total arclength of the contour. Therefore, continuity of the surface and tangent require the constraints $z(0) = 0$, $z(1)=Z$, $r(0) = r(1) = 0$, $\psi(0) = 0$, and $\psi(1)= \pi$.

All the parameters of our model are determined independently, allowing us to numerically calculate the force-extension curves as a function of membrane diameter (Fig.~\ref{M13fig}).  For the three largest membranes, the discrepancy between theory and experiment is comparable to the 20\% variation between different experiments. For the two smallest membranes, the agreement is poor since a plateau region was not observed in the experiments. In this regime, the small extension Hooke's law regime seems to directly transition to the ``overstretching'' regime, which is not treated by our theory. Furthermore, since our effective model has no fitting parameters, one should not expect uniformly good quantitative agreement across a range of membrane sizes. Most of these parameters are only well-defined when the half-micron twist penetration depth is significantly smaller than the membrane size. Numerical results showed quantitative agreement with experimentally measured force and pitch for two differently sized $fd$-$wt$ membranes with lower edge tension (Fig.~\ref{fig4}). Here, the agreement between theory and experiment for the force-extension curve is better due to the presence of plateaus. The agreement for the pitch-extension curve is also reasonable once the extension is large enough for an accurate measure of the twist. We note that a nonzero value of $c^*$ was required for visible twisting, implying that chirality is essential for producing ribbons in the range of applied extensions. 

To summarize, in the absence of external force disk-shaped colloidal membranes are stabilized by the line tension and edge bending energy, despite the tendency of the chirality and the positive $\bar\kappa$ to induce twisted shapes. We have shown that subjecting such membranes to extension yields twisted ribbons. The predictions of the proposed effective theory are in semi-quantitative agreement with the experimental measurements. The quantitative discrepancies occur in the regime where the theory is expected to break down. Future theoretical work should account more explicitly for the liquid crystalline degrees of freedom of the constituent rods and remove the constraint of fixed area. Experimentally, it has been demonstrated that lowering the edge tension by increasing the chirality of the constituent rods spontaneously transforms disk-like membranes into twisted ribbons~\cite{gibaud2012reconfigurable}. The methods developed here could be used to map the free energy landscape associated with such  morphological transitions.

\begin{acknowledgments} This work was supported in part by the National Science Foundation through Grants No. MRSEC-1420382 (A.B., Z.D., L.L.J, R.A.P, and T.R.P), BMAT-1609742 (A.B., M.J.Z, and Z.D.) and No. CMMI-1634552 (L.L.J, R.A.P, and T.R.P).
\end{acknowledgments}
 
%

\end{document}